\documentclass{elsart}
\journal{New Astronomy}

\usepackage{natbib}

\usepackage{graphicx}
\usepackage{epsfig}

\usepackage{amssymb}

\newcommand{\alm}{a_{\ell m}}
\newcommand{\Ylm}{Y_{\ell m}}

\begin{document}

\begin{frontmatter}

 \title{Mysteries on Universe's Largest Observable Scales}


\author{Dragan Huterer$^1$}
\address{$^1$Kavli Institute for Cosmological Physics and Department of Astronomy
and Astrophysics, University of Chicago, Chicago, IL 60637\\
E-mail: dhuterer@kicp.uchicago.edu}

\begin{abstract}
We review recent findings that the universe on its largest scales shows hints
of the violation of statistical isotropy, in particular alignment with the
geometry and direction of motion of the solar system, and missing power at
scales greater than 60 degrees. We present the evidence, attempts to explain it
using astrophysical, cosmological or instrumental mechanisms, and prospects for
future understanding.
\end{abstract}


\end{frontmatter}

\section{Introduction}

The cosmological principle states that the universe is homogeneous and
isotropic on its largest scales. The principle, introduced at the beginning of
any cosmology course, is a crucial ingredient in obtaining most important
results in quantitative cosmology.  For example, assuming the cosmological
principle, cosmic microwave background (CMB) temperature fluctuations in
different directions on the sky can be averaged out, leading to accurate
constraints on cosmological parameters that we have today.  However, there is
no fundamental reason why statistical isotropy must be obeyed by our
universe. Therefore, testing the cosmological principle is one of the crucial
goals of modern cosmology.

Statistical isotropy has only begun to be precision tested recently, with the
advent of first large-scale maps of the cosmic microwave background anisotropy
and galaxy surveys.  Extraordinary full-sky maps produced by the Wilkinson
Microwave Anisotropy Probe (WMAP) experiment, in particular, are
revolutionizing our ability to test the isotropy of the universe on its largest
scales \cite{Bennett_2003, Spergel_2003,Hinshaw_2003,Spergel_2006}.  Stakes are
set even higher with the recent discovery of dark energy that makes the
universe undergo accelerated expansion. It is known that dark energy can
affect the largest scales of the universe --- for example, the clustering scale of
dark energy may be about the horizon size today. Similarly, inflationary models
can induce observable effects on the largest scales via either explicit or
spontaneous  violations of statistical isotropy.  

\section{Multipole vectors}
\label{sec:pol}

Multipole vectors are a new basis that describes the CMB anisotropy (or more
generally, any scalar function on the sky) and are particularly useful in
performing tests of isotropy and alignments.  CMB temperature is traditionally
expressed in harmonic basis, using the spherical
harmonics $\Ylm$. Copi, Huterer \& Starkman (2003; \cite{CHS}) have introduced
an alternative representation in terms of unit vectors
\begin{equation}
  T_{\ell} \equiv \sum_m \alm\Ylm(\theta, \phi)\approx 
  A^{(\ell)} \prod_{i=1}^{\ell}({\hat v}^{(\ell,i)}\cdot {\hat e}) \, ,
\label{eq:vectors}
\end{equation}
where ${\hat v}^{(\ell,i)}$ is the $i^{\rm th}$ multipole vector of the
$\ell^{\rm th}$ multipole. (In fact the right hand side contains terms with
``angular momentum'' $\ell-2$, $\ell-4$ etc.; these are subtracted by taking
the appropriate traceless symmetric combination as described in \cite{CHS} and
\cite{CHSS}.)  In more technical language, Eq.~(\ref{eq:vectors}) states the
equivalence between a symmetric, traceless tensor of rank $\ell$ (middle term)
and the outer product of $\ell$ unit vectors (last term).  Note that the signs
of all vectors can be absorbed into the sign of $A^{(\ell)}$.  This
representation is unique, and the right-hand side contains the familiar
$2\ell+1$ degrees of freedom -- two dof for each vector, plus one for
$A^{(\ell)}$.

An efficient algorithm to compute the multipole vectors has been presented in
\cite{CHS} and is publicly available \cite{MV_code}; other algorithms have
been proposed as well \cite{Katz2004,Weeks04,Helling}. Interestingly, after the
publication of the CHS paper \cite{CHS}, Weeks \cite{Weeks04} pointed out that
multipole vectors have actually first been invented by Maxwell \cite{Maxwell}
more than 100 years ago!

The relation between multipole vectors and the usual harmonic basis is very
much the same as that between cartesian and spherical coordinates of standard
geometry: both are complete bases, but specific problems are much more easily
solved in one basis than the other. In particular, we and others have
found that multipole vectors are particularly well suited for tests of planarity of
the CMB anisotropy pattern. Moreover, a number of interesting theoretical
results have been found; for example, Dennis \cite{Dennis2005} analytically computed the
two-point correlation function of multipole vectors for a gaussian random, isotropic 
underlying field, while in Copi et al. \cite{CHSS} we have studied the relation of multipole
vectors to maximum angular momentum dispersion axes, maxima/minima directions,
and other related quantities that have been proposed to study the CMB.

\section{Alignments with the Solar System}
\label{sec:align}

Armed with this new representation of the CMB anisotropy, we have set out to
study the morphology of CMB anisotropies on large angular scales.  Prior to our
work, Tegmark et al.\ \cite{TOH} found that the octopole is planar and that the
quadrupole and octopole planes are aligned. In Schwarz et al.\ \cite{SSHC}, we
have investigated the quadrupole-octopole shape and orientation using the
multipole vectors.  Quadrupole defines two vectors and therefore one ``oriented
area'' vector $\hat w^{(\ell;i,j)}\equiv v^{(\ell, i)}\times v^{(\ell,
j)}$. Octopole defines three multipole vectors and therefore three
normals. Hence there are a total of four planes determined by the quadrupole
and octopole.

\begin{figure}[!t]
\begin{center}
\psfig{file=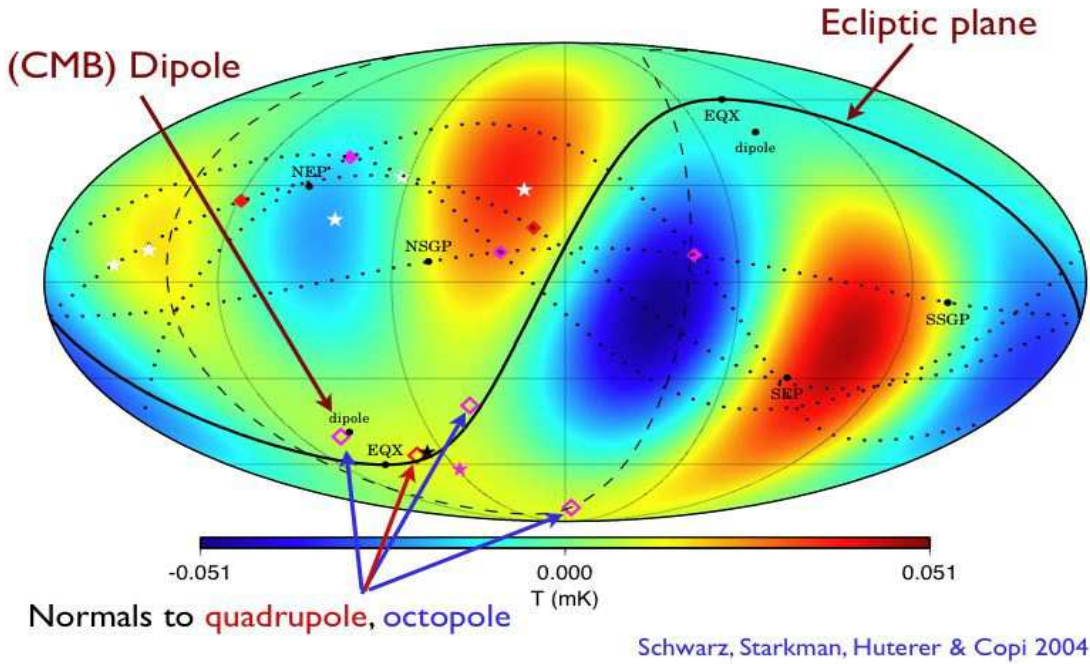,height=4.1in,width=5.95in, angle=0}
\caption{Quadrupole and octopole ($\ell=2$ and $3$) of the WMAP sky map in
galactic coordinates, shown with the ecliptic plane, the supergalactic plane
(SGP), the equinoxes and the cosmological dipole.  We also show the four
normals to the planes defined by vectors that describe the quadrupole and
octopole temperature anisotropy; one normal is defined by the quadrupole and
three by the octopole. Note that three out of four
normals lie very close to both the equinoxes and the dipole direction. The
probability of these alignments being accidental is about one part in a
thousand. Moreover, the ecliptic plane traces out a locus of zero of
the combined quadrupole and octopole over a broad swath of the sky ---- neatly
separating a hot spot in the northern sky from a cold spot in the south. These
apparent correlations with the solar system geometry are puzzling
and currently unexplained.  }
\label{fig:align}
\end{center}
\end{figure}

In Schwarz et al.\ we found that

\begin{itemize}
\item the normals to these four planes are aligned with the direction of the
cosmological dipole and with the equinoxes at a level inconsistent with
Gaussian random, statistically isotropic skies at $99.95$\% C.L.;

\item the quadrupole and octopole planes are orthogonal to the ecliptic
at the $98.5$\% C.L.;

\item the ecliptic threads between a hot and a cold spot of the combined
quadrupole and octopole map, following a node line across about $1/3$ of the
sky and separating the three strong extrema from the three weak extrema of the
map; this is unlikely at about the $95$\% C.L.;

\item the four area vectors of the quadrupole and octopole are mutually close
(i.e.\ the quadrupole and octopole planes are aligned) at the $99.9$\% C.L.
\end{itemize}

(These numbers refer to the TOH map; other maps give similar results as Table 3
of Ref.~\cite{CHSS} shows.)  While not all of these alignments are
statistically independent, they are clearly surprising, highly statistically
significant (at $>99.9$\% C.L.), and unexpected in the standard inflationary
theory and the accepted cosmological model.

Particularly puzzling are the alignments with the solar system features.  CMB
anisotropy should clearly not be correlated with our local habitat. While the
observed correlations seem to hint that there is contamination by a foreground
or perhaps scanning strategy of the telescope, closer inspection (see the next
two Sections) reveals that there is no one obvious way to  explain the observed
correlations.

Our studies (see \cite{CHSS}) indicate that the observed alignments are
equinoctic/ecliptic ones (and/or correlated with the dipole direction), and
{\it not} alignments with the Galactic plane: the alignments of the quadrupole
and octopole planes with the equinox/ecliptic/dipole directions are more significant
than those for the Galactic plane.  This conclusion is supported by the
foreground analysis (see the next Section).  Moreover, it is remarkably curious
that it is precisely the ecliptic alignment that has been found on somewhat
smaller scales using the power spectrum analyses of statistical isotropy
\cite{NS_asymmetry, Bernui}.

Finally, it is important to make sure that the results are unbiased due to unfairly
chosen statistics. We have studied this issue extensively in \cite{CHSS}. Two
natural choices of statistics which define ordering relations on the three
dot-products between the quadrupole and octopole area vectors $A_i$, each lying
in the interval $[0,1]$, are:
\begin{eqnarray}
S &\equiv& {1\over 3} \left (A_1 + A_2 + A_3\right ), \,\,\,\,{\rm and}\nonumber \\
T &\equiv& 1-{1\over 3}\left [(1-A_1)^2 + (1-A_2)^2 + (1-A_3)^2\right ].
\end{eqnarray}
Both $S$ and $T$ can be viewed as the suitably defined ``distance'' to the
vertex $(A_1, A_2, A_3)=(0, 0, 0)$.  A third obvious choice,
$(A_1^2+A_2^2+A_3^2)/3$, is just $2S-T$. To test alignment of the quadrupole
and octopole planes (or associated area vectors) we quoted
the $S$ statistic numbers; $T$ gives similar results. 

To test alignments of multipole planes, we define the plane as the one whose
normal, $\hat{n}$, has the largest dot product with the sum of the area vectors
\cite{CHSS}. Since $\vec{w}_i \cdot \hat{n}$ is defined only up to a sign ---
$\vec{w}_i$ is headless --- we take the absolute value of each dot
product. Therefore, we find $\hat{n}$ that maximizes
\begin{equation}
\mathcal{S}\equiv \frac 1 {N_{\ell}} 
\sum_{i=1}^{N_\ell} \left |\vec{w}_i \cdot \hat{n}\right |,
\label{eq:S_single_mult}
\end{equation}
where $N_\ell$ is the total number of area vectors considered. Alternatively,
generalizing the definition in \cite{TOH}, one can find the direction that
maximizes the angular momentum and compare the maximal angular momentum (for
the quadrupole plus octopole) with that from simulated isotropic skies  \cite{CHSS}
\begin{equation}
 \hat L^2_\ell \equiv { \sum_{m=-\ell}^\ell m^2 \vert a_{\ell m}\vert^2 \over 
\ell^2 \sum_{m=-\ell}^\ell \left| \alm \right|^2  },
\label{eqn:angmomn}
\end{equation}
which gives similar results as the $\mathcal{S}$ statistic for the alignment
of $\ell=2$ and $\ell=3$.

\section{Foregrounds}
\label{sec:foregr}

While the statistical significance of the observed vs.\ the Galactic alignments
of the quadrupole and octopole, by itself, may not not sufficient to rule out
the Galactic contamination, we have explored several lines of reasoning which
suggest that Galactic foregrounds are not the cause of the alignments (see also
studies by \cite{SS,Bielewicz05,Abramo}).

\begin{figure}[!t]
\begin{center}
  \includegraphics[width=2.1in,angle=-90]{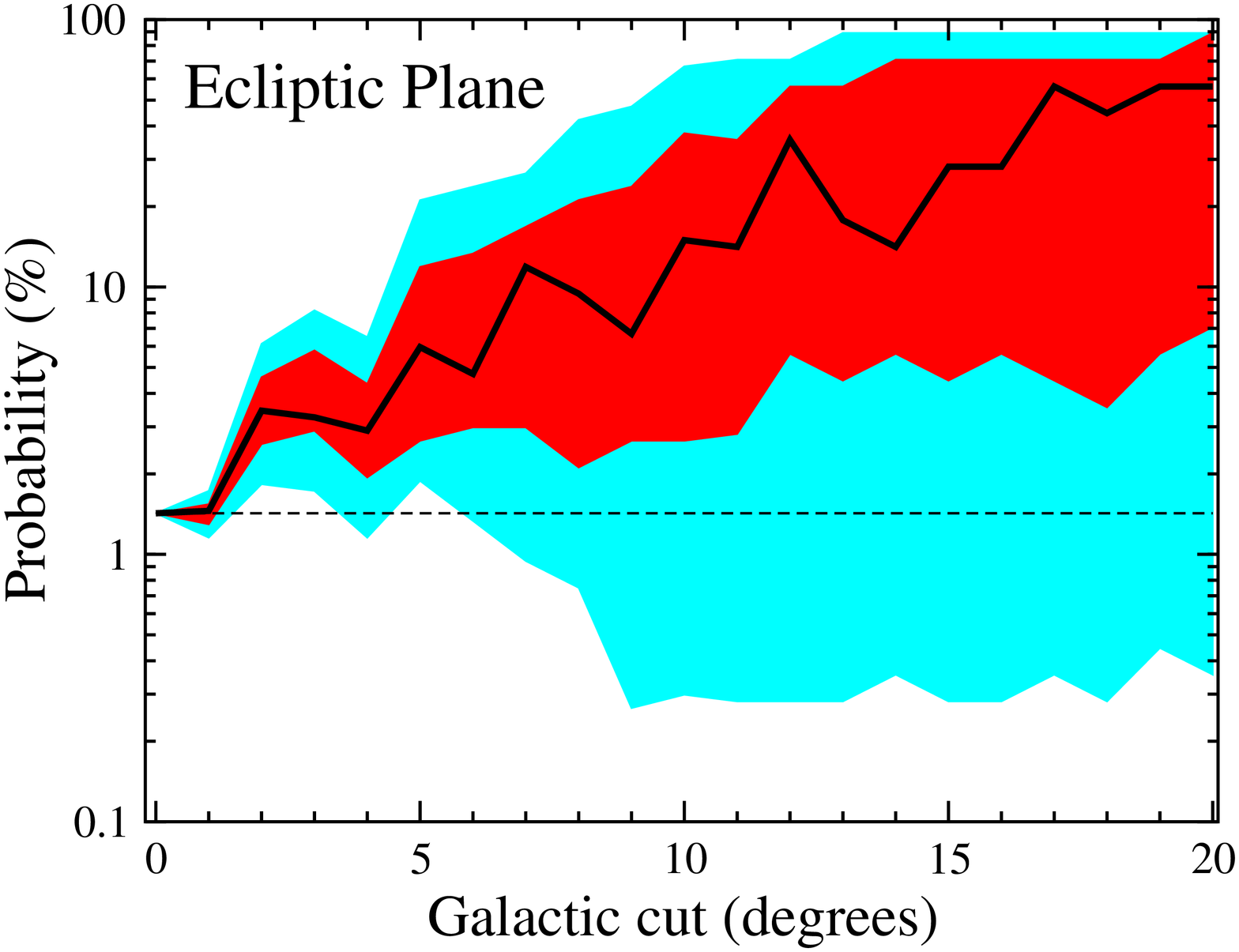}\hspace{-0.2cm}
  \includegraphics[width=2.1in,angle=-90]{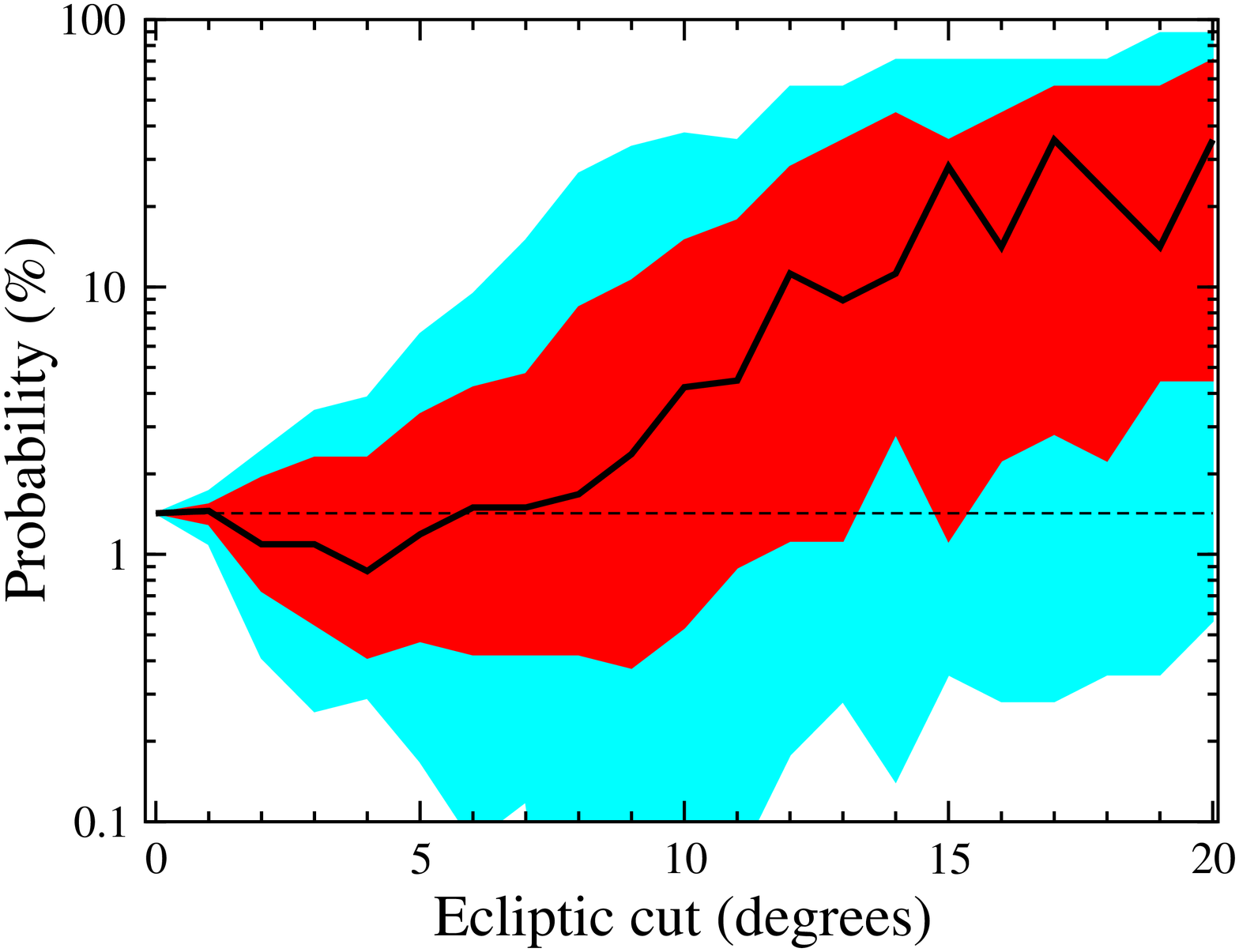}
  \caption{Quadrupole-octopole probabilities for the TOH map for an
    increasingly larger isolatitude cut of $\pm$(degrees shown), performed
    symmetrically around the Galactic plane (left panel) or the ecliptic plane
    (right panel). We consider the $S$ statistic probabilities applied to the
    ecliptic plane alignment --- the solid line is the mean value, while the dark and
    light regions represent 68\% C.L.\ and 95\% C.L.\ regions, respectively,
    from 1000 realizations of reconstructed $a_{\ell m}$ coefficients. The
    dashed line denotes the probability obtained from the full-sky map,
    corresponding to the case of zero cut. Adopted from \cite{CHSS}. }
\label{fig:cutsky}
\end{center}
\end{figure}

First, we have tried adding (or subtracting) known, measured Galactic
contamination to WMAP maps and observing how the multipole vectors move
\cite{CHSS}. In the large-foreground limit, the quadrupole vectors move near
the $z$-axis and the normal into the Galactic plane, while for the octopole all
three normals become close to the Galactic disk at $90^\circ$ from the Galactic
center. Therefore, as expected Galactic foregrounds lead to Galactic, and not
ecliptic, correlations of the quadrupole and octopole.

Second, in \cite{CHSS}, we have shown that the known Galactic foregrounds
possess a multipole vector structure very different from that of the observed
quadrupole and octopole. The quadrupole is nearly pure $Y_{22}$ in
the frame where the $z$-axis is parallel to the dipole (or $\hat w^{(2,1,2)}$
or any nearly equivalent direction), while the octopole is dominantly $Y_{33}$
in the same frame.  Mechanisms which produce an alteration of the microwave
signal from a relatively small patch of sky --- and all of the recent proposals
fall into this class --- are most likely to produce aligned $Y_{20}$ and
$Y_{30}$ (essentially because the multipole vectors of the affected multipoles
will all be parallel to each other, leading to a $Y_{\ell 0}$ in this frame).

Most of the results discussed so far have been obtained using reconstructed
full-sky maps of the WMAP observations \cite{Bennett_2003,TOH, LILC}.  Results
with the reconstructed full-sky map in the presence of the sky cut is shown in
Fig.~(\ref{fig:cutsky}): even with a cut of a few degrees (iso-latitude, for
simplicity), the errors in the reconstructed anisotropy pattern, and the
directions of multipole vectors, are too large to allow drawing quantitative
conclusions about the observed alignments. Figure \ref{fig:cutsky} does show,
however, that the cut-sky alignment probabilities, while very uncertain, are
consistent with the full-sky values. Ultimately, one will want to check for the
low-$\ell$ alignments on Markov chain Monte Carlo maps, where realizations of
the reconstructed the anisotropy pattern over the whole sky are based on the
observations outside of the Galactic cut. While in principle straightforward
(see e.g.~\cite{ODwyer}), the key issue in this approach that requires
considerable care is modeling of the foregrounds.

\section{Quest for an explanation}

Understanding the origin of CMB anomalies is clearly important, as the observed
alignments of power at large scales are inconsistent with predictions of
standard cosmological theory.  A number of authors have attempted to explain
the observed quadrupole-octopole correlations in terms of a new foreground --- 
for example the Rees-Sciama effect \cite{Rakic06}, interstellar dust \cite{Frisch05},
local voids \cite{Inoue06},\cite{Ghosh}. Most if not all of these proposals have a difficult
time explaining the anomalies without severe fine tuning.  For example, Vale \cite{Vale05}
cleverly suggested that the moving lens effect, with the Great Attractor as a source,
might be responsible for the extra anisotropy; however Cooray \& Seto
\cite{Cooray05} have argued that the lensing effect is far too small and requires too
large a mass of the Attractor.  

In Gordon et al.\ \cite{Gordon} we have explored the
alignment mechanisms in detail, and studied additive models where the
temperature is added to the intrinsic temperature
\begin{equation}
T_{\rm observed}(\hat{\bf n})=T_{\rm intrinsic}(\hat{\bf n})+T_{\rm add}(\hat{\bf n})
\end{equation}
where $T_{\rm add}(\hat{\bf n})$ is the additive term -- perhaps contamination
by a foreground, perhaps an additive instrumental or cosmological effect. We
have shown that additive modulations of the CMB sky that ameliorate the
alignment problems tend to {\it worsen} the overall likelihood at large scales
(they still may pick up positive likelihood contribution from higher
multipoles).  The intuitive reason for this is that there are two penalties
incurred by the additive modulation. First, since the temperature at large
scales is lower than expected, one typically needs to arrange for an accidental
cancellation between $T_{\rm intrinsic}$ and $T_{\rm add}$. Second, certain
$\alm$ in the dipole frame are observed to be suppressed relative to the
expectation (see Table I in Ref.~\cite{Gordon}) -- but zero is actually the
most likely value of any given $\alm$, so likelihood with $T_{\rm intrinsic}$
is again penalized.

\begin{figure}[!t]
\begin{center}
\epsfig{file=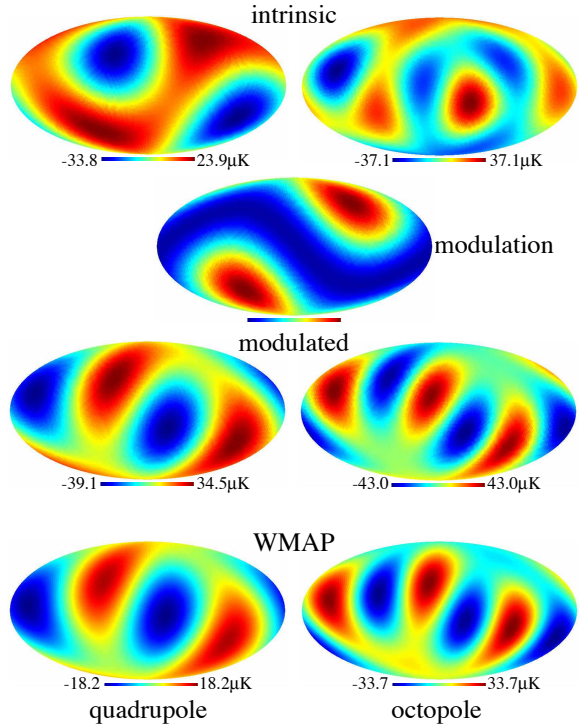, width=3in}
\caption{A realization of the multiplicative model where the quadrupole (left
   column) and octopole (right column) exhibit an alignment similar to WMAP.
   First row: intrinsic (unmodulated) sky from a Gaussian random isotropic
   realization.  Second row (single column): the quadrupolar modulation $
   \propto -[1-7 Y_{20}(\hat{\bf n})]$ in the dipole direction.  Third row: the
   modulated sky of the observed CMB.  Fourth row: WMAP full-sky quadrupole and
   octopole.  }
\end{center}
\end{figure} 

Instead, the multiplicative mechanisms, where the intrinsic temperature is
multiplied by a spatially varying modulation, are more promising. As a proof of
principle, we suggested a toy-model modulation
\begin{equation}
T_{\rm observed}(\hat{\bf n})=f\left [1+w_2Y_{20}(\hat{\bf n})\right ]\,
T_{\rm intrinsic}(\hat{\bf n}),
\end{equation}
(where the modulation is a pure $Y_{20}$ along the dipole axis), we have shown
that the likelihood of the WMAP data can be increased by a factor of
$\exp{(16/2)}$ and, at the same time, the probability of obtaining a sky with
more alignment (e.g.\ higher angular momentum statistic) is increased 200 times, to
45\%. (Spergel et al.\ 2006 \cite{Spergel_2006} thereafter did a similar study, 
generalizing the multiplicative modulation to eight free parameters corresponding
to all components of the dipole and quadrupole and finding the highest likelihood fit;
see their Fig.~(26)).

Finally we have considered a possibility of an imperfect instrument,
where the instrumental response to the signal $T(\hat{\bf n})$ is nonlinear
\begin{equation}
T_{\rm observed}(\hat{\bf n})=T(\hat{\bf n})+\alpha_2\,T(\hat{\bf n})^2
+ \alpha_3\,T(\hat{\bf n})^3  + \ldots
\end{equation}
Since the biggest signal on the sky is the dipole (of order mK), leakage of
about 1\% (i.e.\ $\alpha_1\approx\alpha_2\approx 0.01$), if judiciously chosen,
can produce the quadrupole and octopole that are as observed {\it and are aligned
with the dipole}. Unfortunately (or fortunately!), WMAP detectors are known to
be linear to much better than 1\%, so this particular realization of the
instrumental explanation does not work. As an aside, note that this type of
explanation needs to assure that the higher multipoles are not aligned with the
dipole/ecliptic, and moreover, requires essentially no intrinsic power at large
scales (that is, even less than what is observed).

\section{Missing angular power at large scales}

Spergel et al.\ \cite{Spergel_2003} have found that the two point correlation function,
$C(\hat{\bf n}\cdot \hat{\bf n}')\equiv \langle T(\hat{\bf n}) T(\hat{\bf n}')\rangle$,
nearly vanishes on scales greater than about 60 degrees, contrary  to what
the standard $\Lambda$CDM theory predicts, and in agreement with the same
finding obtained from COBE data about a decade earlier \cite{DMR4}.
Using the statistic
\begin{equation}
S_{1/2} \equiv \int_{-1}^{1/2} \left[ C(\theta)\right]^2 d (\cos\theta).
\label{eqn:Shalf}
\end{equation}
Spergel et al.\ found that only $0.15\%$ of the elements in their Markov
chain of $\Lambda$CDM model CMB skies had lower values of $S_{1/2}$ than the
observed sky.

\begin{figure*}[!t]
\includegraphics[width=2.8in,height=3.1in,angle=-90]{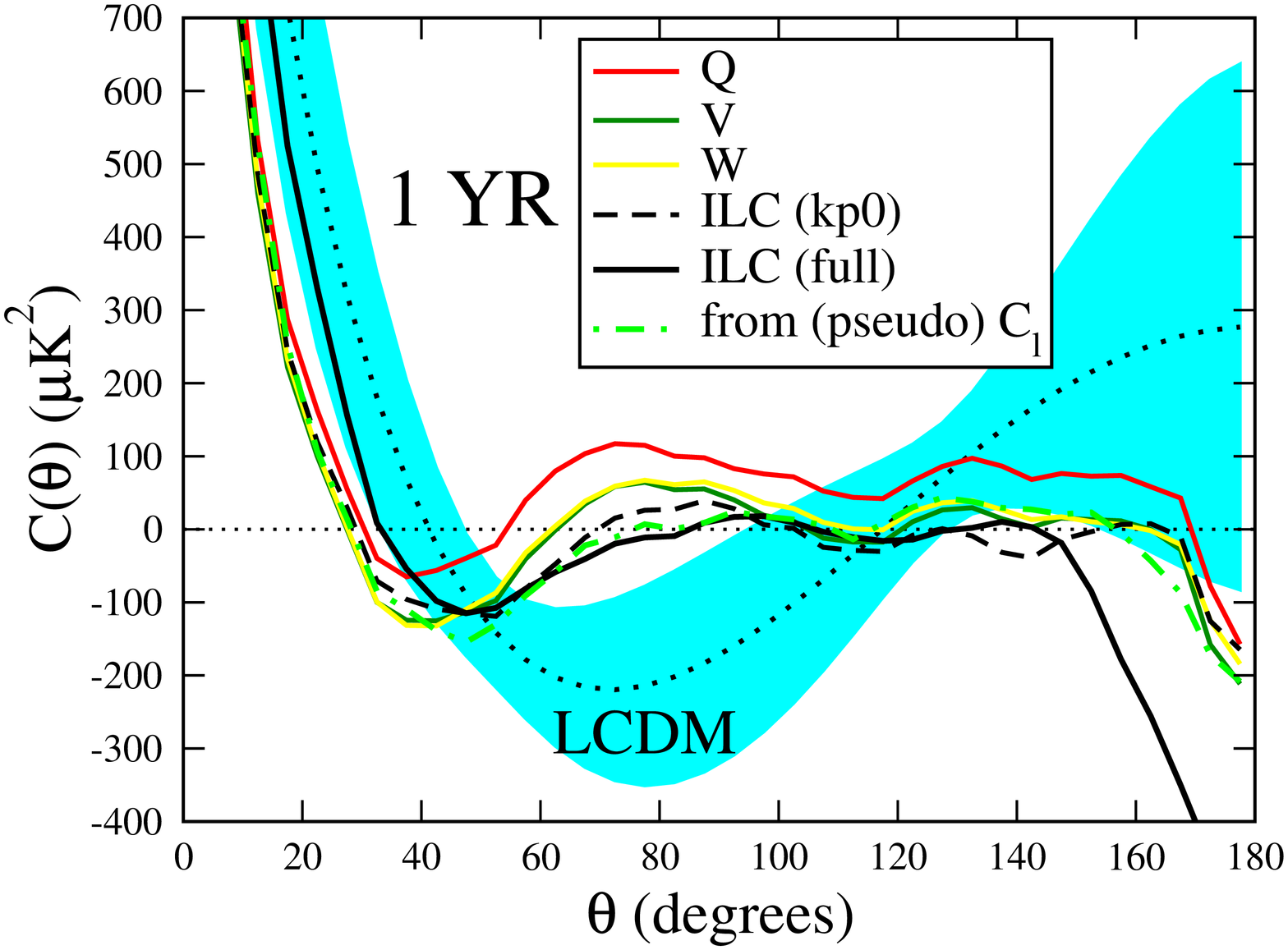}\hspace{-0.75cm}
\includegraphics[width=2.8in,height=3.1in,angle=-90]{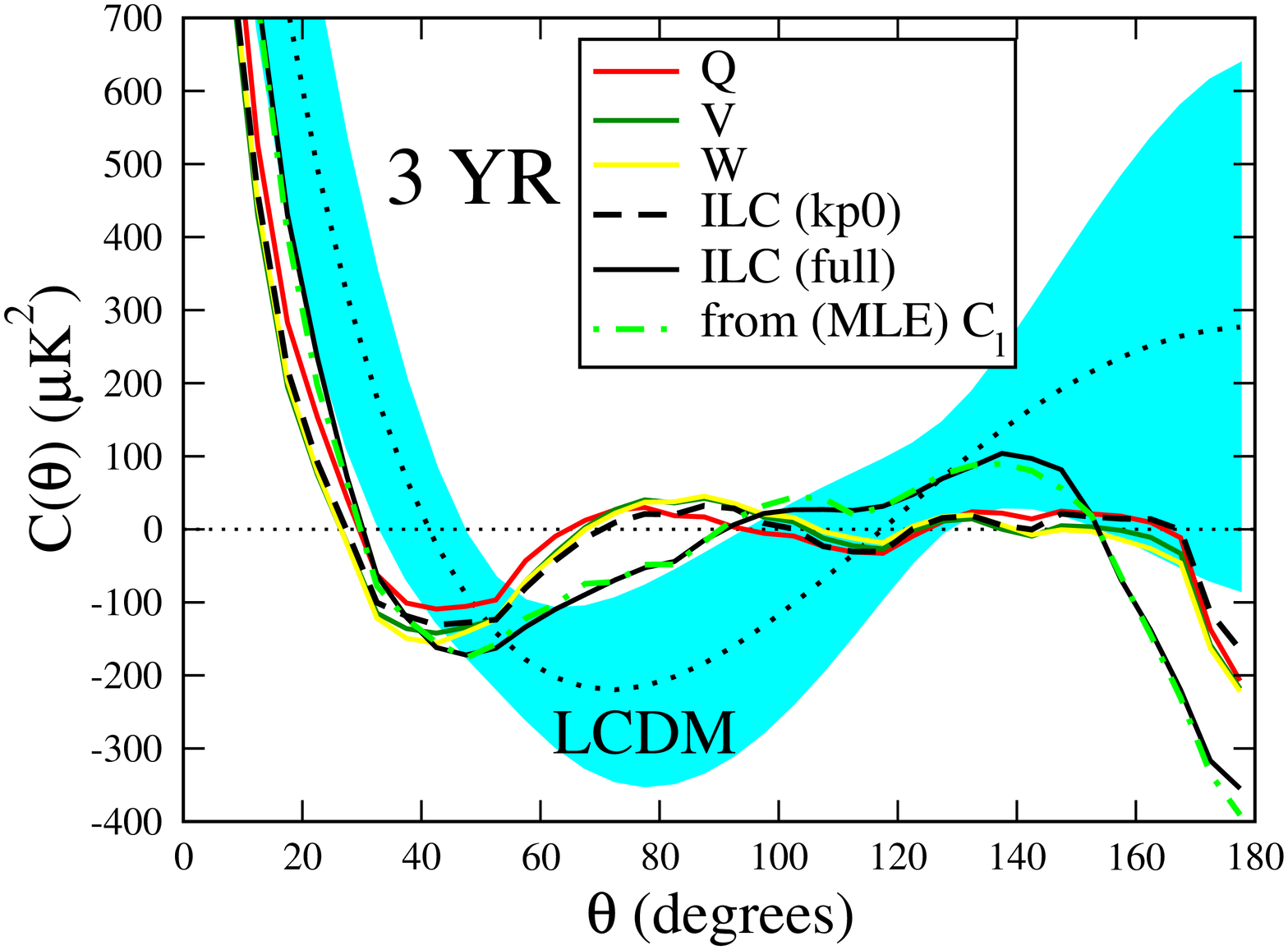}
\caption{Two point angular correlation function, ${\mathcal C}(\theta) \equiv
{\overline{ T({\hat e}_1)T({\hat e}_2)}}_\theta$ computed in pixel space, for
three different bands masked with the kp0 mask. Also shown is the correlation
function for the ILC map with and without the mask, and the value expected for
a statistically isotropic sky with best-fit $\Lambda$CDM cosmology together
with 68\% error bars.  Left panel: year 1 results.  Right panel: year 123
results. Even by eye, it is apparent that masked year 123 maps have $C(\theta)$
that is consistent with zero at $\theta\gtrsim 60$ deg, even more so than in
year 1 maps (at the $0.03$\%-$0.15$\% level depending on the map used). We also
show the $C(\theta)$ computed from the ``official'' published $C_\ell$, which
(at $\ell<10$) are the pseudo-$C_\ell$ in year 1, the and MLE $C_\ell$ in year
123. Clearly, the maximum likelihood estimator (MLE)-based $C_\ell$, as well as
${\mathcal C}(\theta)$ computed from the full-sky ILC maps, are in significant
disagreement with the angular correlation function computed from cut-sky maps.}
\label{fig:ctheta}
\end{figure*}

We have revisited the angular two point function in the 3-yr WMAP data in
Ref.~\cite{Ctheta_us}. We found that the two-point function computed from the
various cut-sky maps shows an even stronger lack of power, now significant at
the $0.03$\%-$0.15$\% level depending on the map used; see
Fig.~(\ref{fig:ctheta}). However, we also found that, while $C(\theta)$ computed
in pixel space over the unmasked sky agrees with the harmonic space calculation
that uses the pseudo-$C_\ell$ estimator, it disagrees with the $C_\ell$
obtained using the maximum likelihood estimator (advocated in the 3rd year WMAP
release \cite{Spergel_2006}). The MLE-based $C_\ell$ lead to $C(\theta)$
that is low (according to the $S_{1/2}$ statistic) only at the 8\% level.  This
is illustrated in the right panel of Fig.~(\ref{fig:ctheta}).  
We are concerned that the full-sky maximum-likelihood map making algorithm is
inserting significant extra large angle power into precisely those portions of
the sky where we have the least reliable information. Clearly, the definitive judgment
of the large-angle power has not yet been made. 

Finally, here we note that the vanishing of power is much more apparent in real
space (as in $C(\theta)$) than in multipole space (as in $C_\ell$).  The
harmonic-space quadrupole and octopole are only moderately low (e.g.\
\cite{ODwyer}), and it is really a range of low multipoles that conspire to
make up the vanishing $C(\theta)$.  Therefore, theoretical efforts to explain
``low power on large scales'' should focus to explain the low
$C(\theta)$ at $\theta\gtrsim 60$ deg.

\section{Discussion and  Future Prospects}
\label{sec:discussion}

If indeed the observed $\ell=2$ and $3$ CMB fluctuations are not cosmological,
one must reconsider all CMB results that rely on low $\ell$s, such as the
normalization, $A$, of the primordial fluctuations and any constraint on the
running $d n_s/d\log{k}$ of the spectral index of scalar
perturbations. Moreover, the CMB-galaxy cross-correlation, which has been used
to provide evidence for the Integrated Sachs-Wolfe effect and hence the
existence of dark energy, also gets contributions from the lowest multipoles
(though the main contribution comes from slightly smaller scales, $\ell\sim
10$). Finally, it is quite possible that the underlying physical mechanism does
not cut off abruptly at the octopole, but rather affects the higher multipoles.
Indeed, several pieces of evidence have been presented for anomalies at $l>3$
\cite{LandMagueijoAoE,NS_asymmetry}; see also relevant work in
Refs.~\cite{Dore03,Chiang,Vielva,Freeman,McEwen,LandMagueijo_other,Hajian}.

So far no convincing explanation has been offered. In fact, a no-go argument
has been given by Gordon et al.~\cite{Gordon}, reasoning that additive
mechanisms for adjusting the intrinsic CMB anisotropy lead to a {\it lower}
likelihood at low $\ell$ than the observed sky. Therefore, it appears that a
multiplicative mechanism is at work, whether it is astrophysical, instrumental
or cosmological. 

While the further WMAP data (4-year, 8-year etc) is not expected to change any
of the observed results, our understanding and analysis techniques are likely
to improve. Much work remains to study the large-scale correlations using
improved foreground treatment, accounting even for the subtle systematics, and
in particular studying the time-ordered data from the spacecraft. The Planck
experiment will be of great importance, as it will provide maps of the largest
scales obtained using a very different experimental approach than WMAP ---
measuring the absolute temperature rather than temperature
differences. Polarization maps, when and if available at high enough
signal-to-noise at large scales (which may not be soon), will be a fantastic
independent test of the alignments, and in particular each explanation for the
alignments, in principle, also predicts the statistics of the polarization pattern on the sky.

The quest for an answer has whetted the appetite of cosmologists to understand
the structure of the universe on its largest scales.

\bigskip
{\it Acknowledgments:} Topics discussed in these proceedings are a
product of collaborative projects with Craig Copi, Dominik Schwarz, Glenn
Starkman, Chris Gordon, Wayne Hu and Tom Crawford.  The author has been
supported by the NSF Astronomy and Astrophysics Postdoctoral Fellowship under
Grant No.\ 0401066.


\begin{thebibliography}{99}
\frenchspacing

\bibitem{Bennett_2003}
       C.~L.~Bennett et.\ al., Astrophys.\ J.\ {\bf 148}, S1 (2003),
       astro-ph/0302207.
\bibitem{Spergel_2003}
       D.~N.~Spergel et al.\, Astrophys.\ J.\ {\bf 148}, S175 (2003),
       astro-ph/0302209.
\bibitem{Hinshaw_2003}
       G.~Hinshaw et al.\, Astrophys.\ J.\ {\bf 148}, S135 (2003),
       astro-ph/0302217.
\bibitem{Spergel_2006}
       D.~N.~Spergel et al.\,   astro-ph/0603449.
\bibitem{CHS}
       C.~J.~Copi, D.~Huterer, and G.~D.~Starkman, 
       Phys.\ Rev.\ {\bf D70}, 043515 (2004), 
       astro-ph/0310511.
\bibitem{CHSS}
       C.~J.~Copi, D.~Huterer, D.~J.~Schwarz, and G.~D.~Starkman, 
       Mon.\ Not.\ Roy.\ Astron.\ Soc.\ {\bf 367}, 79 (2006),
       astro-ph/0508047.
\bibitem{MV_code}
       http://www.phys.cwru.edu/projects/mpvectors/
\bibitem{Katz2004}
       G.~Katz G. and J.~Weeks, 
       Phys.\ Rev.\ {\bf D70}, 063527 (2004),
       astro-ph/0405631.
\bibitem{Weeks04}
       J.R.~Weeks, astro-ph/0412231.
\bibitem{Helling}
       R.C.~Helling, P.~Schupp and T. Tesileanu,
       astro-ph/0603594.
\bibitem{Maxwell}
       J.C.\ Maxwell, ``A Treatise on Electricity  and Magnetism'', Clarendon Press, London, 1891
\bibitem{Dennis2005}
        M.~R.~Dennis,
        J. Phys. A: Math. Gen. {\bf 38}, 1653 (2005).
\bibitem{TOH}
       M.~Tegmark, A.~de Oliveira-Costa and A.~J.~S.~Hamilton, 
       Phys.\ Rev.\ D {\bf 68}, 123523 (2003),
       astro-ph/0302496.
\bibitem{SSHC}
       D.~J.~Schwarz, G.~D.~Starkman, D.~Huterer, and C.~J.~Copi, 
       Phys.\ Rev.\ Lett.\ {\bf 93}, 221301 (2004),
       astro-ph/0403353.
\bibitem{NS_asymmetry}
       H.~K.~Eriksen, F.~K.~Hansen, A.~J.~Banday, K.~M.~Gorski, and P.~B.~Lilje,
       Astrophys.\ J.\  {\bf 605}, 14 (2004);
       {\bf 609}, 1198 (2004) [Erratum], astro-ph/0307507;
       F.K.\ Hansen, A.J.\ Banday and K.M.\ Gorski, 
       Mon.\ Not.\ Roy.\ Astron.\ Soc.\ {\bf 354}, 641 (2004).
\bibitem{Bernui}
       A. Bernui, B.\ Mota,  M.J.\ Reboucas, and R.\ Tavakol, astro-ph/0511666;
       A.~Bernui et al., astro-ph/0601593
\bibitem{SS}
       A.~Slosar and U.~Seljak, 
       Phys.\ Rev.\ {\bf D70}, 083002 (2004),
       astro-ph/0404567.
\bibitem{Bielewicz05}
       P.~Bielewicz, H.K.~Eriksen, A.J.~Banday, K.M.~G{\' o}rski and P.B.~Lilje,
       Astrophys.\ J.\ {\bf 635}, 750 (2005), astro-ph/0507186
\bibitem{Abramo}
       L.R.\ Abramo et al., astro-ph/0604346;
       L.R.\ Abramo, S.\ Sodre and A.\ Wuensche, astro-ph/0605269.
\bibitem{LILC}
       H.~K.~Eriksen, A.~J.~Banday, K.~M.~Gorski, and P.~B.~Lilje,
       Astrophys.\ J.\  {\bf 612}, 633 (2004),  astro-ph/0403098.
\bibitem{ODwyer}
       I.J.\ O'Dwyer et al., Astrophys.\ J.\  {\bf 617}, L99 (2004),
       astro-ph/0407027.
\bibitem{LandMagueijoAoE}
       K.~Land and J.~Magueijo,
       Mon.\ Not.\ Roy.\ Astron.\ Soc.\ {\bf 362}, L16 (2005), 
       astro-ph/0407081; 
       Phys.~Rev.~Lett.\ {\bf 95}, 071301 (2005), 
       astro-ph/0502237; 
\bibitem{Rakic06}
       A.~Raki\'c, S.~R\"as\"anen and D.~J.~Schwarz,
       Mon. Not.\ Roy.\ Astron.\ Soc.\ {\bf 369}, L27 (2006),  
       astro-ph/0601445.
\bibitem{Frisch05}
       P.~C.~Frisch, Astrophys.\ J.\  {\bf 632}, L143 (2005),
       astro-ph/0506293.
\bibitem{Inoue06}
       K.~T.~Inoue and J.~Silk,
       astro-ph/0602478.
\bibitem{Ghosh}
       T.~Ghosh, A.~Hajian and T.~Souradeep,
       astro-ph/0604279.
\bibitem{Vale05}
       C.~Vale, astro-ph/0509039.
\bibitem{Cooray05}
       A.~Cooray and N.~Seto,
       JCAP {\bf 0512}, 004 (2005),
       astro-ph/0510137.
\bibitem{Gordon}
       C.~Gordon, W.~Hu, D.~Huterer and T.~Crawford,
       Phys.\ Rev.\ D {\bf 72}, 103002 (2005), 
       astro-ph/0509301.
\bibitem{DMR4}
       G.~Hinshaw et.\ al., Astrophys.\ J.\ {\bf 464}, L25 (1996),
       astro-ph/9601061.
\bibitem{Ctheta_us}
       C.~J.~Copi, D.~Huterer, D.~J.~Schwarz, and G.~D.~Starkman, 
       Phys.\ Rev.\ D, submitted, astro-ph/0605135.
\bibitem{Dore03}
       O.~Dore, G.~P.~Holder, and A.~Loeb, 
       Astrophys.\ J.\ {\bf 612}, 81 (2004), astro-ph/0309281. 
\bibitem{Chiang}
       L.~Y.~Chiang, P.~D.~Naselsky, O.~V.~Verkhodanov and M.~J.~Way,
       Astrophys.\ J.\ {\bf 590}, L65 (2003),
       astro-ph/0303643;
       P.~Naselsky, L.~Y.~Chiang, P.~Olesen and I.~Novikov,
       Phys.\ Rev.\ D {\bf 72}, 063512 (2005), 
       astro-ph/0505011.
\bibitem{Vielva}
       P.~Vielva, E.~Martinez-Gonzalez, R.~B.~Barreiro, J.~L.~Sanz, and
       L.~Cayon,
       Astrophys.\ J.\  {\bf 609}, 22 (2004), 
       astro-ph/0310273;
       M.~Cruz, E.~Martinez-Gonzalez, P.~Vielva and L.~Cayon,
       Mon.\ Not.\ Roy.\ Astron.\ Soc.\  {\bf 356}, 29 (2005),
       astro-ph/0405341;
       Y.~Wiaux, P.~Vielva, E.~Martinez-Gonzalez and P.~Vandergheynst,
       Phys.~Rev.~Lett.\ {\bf 96}, 151303 (2006), 
       astro-ph/0603367.
\bibitem{Freeman}
        P.~E.~Freeman, C.~R.~Genovese, C.~J.~Miller, R.~C.~Nichol, and 
	L.~Wasserman,  
	Astrophys.\ J.\  {\bf 638}, 1 (2006), astro-ph/0510406.
\bibitem{McEwen}
       J.D.~McEwen, M.P.~Hobson, A.N.~Lasenby and D.J.~Mortlock,
       Mon.\ Not.\ Roy.\ Astron.\ Soc.\ {\bf 359}, 1583 (2005),
       astro-ph/0406604.
\bibitem{LandMagueijo_other}
       K.~Land and J.~Magueijo,
       Mon.\ Not.\ Roy.\ Astron.\ Soc.\ {\bf 362}, 838 (2005), 
       astro-ph/0502574; 
       K.~Land and J.~Magueijo, 
       Phys.\ Rev.\ D {\bf 72}, 101302 (2005),
       astro-ph/0507289. 
\bibitem{Hajian}
       A.\ Hajian, T.\ Souradeep and N.\  Cornish,  Astrophys.\ J.\ {\bf 618}, L63 (2004);
       A.~Hajian and T.~Souradeep, astro-ph/0501001;       
       S.\ Basak, A.~Hajian and T.~Souradeep, Phys.\ Rev.\ D 74, 021301 (2006);
       A.~Hajian and T.~Souradeep, astro-ph/0607153.       
\end{thebibliography}
\end{document}